\documentclass[twocolumn,showpacs,preprintnumbers,prl]{revtex4-1}
\usepackage{graphicx,bm,amsmath,amssymb}

\newcommand{\be}{\begin{equation}}
\newcommand{\ee}{\end{equation}}
\newcommand{\bea}{\begin{eqnarray}}
\newcommand{\eea}{\end{eqnarray}}

\begin{document}

\title{Many-body dynamics of the decay of excitons of different charges in a quantum dot}
\author{J.~A.~Andrade}
\author{A.~A.~Aligia}
\author{Pablo~S.~Cornaglia}
\affiliation{Centro At\'{o}mico Bariloche and Instituto Balseiro, Comisi\'{o}n Nacional
de Energ\'{\i}a At\'{o}mica, CONICET, 8400 Bariloche, Argentina}
\date{\today}

\begin{abstract} 
We calculate the photoluminescence spectrum of a single semiconductor quantum dot strongly coupled to a continuum 
as a function of light frequency, gate voltage, and magnetic field. The spectrum is dominated by the recombination of several 
excitonic states 
which can be considered as
quantum quenchs in which the many-body nature of the system is suddenly changed between 
initial and final states. This is associated with an Anderson orthogonality catastrophe with 
a power-law singularity at the threshold.
We explain the main features observed experimentally in the 
region of stability  of the trion $X^-$, the neutral exciton $X^0$ and the gate voltage induced transition between them.

\end{abstract}

\pacs{78.67.Hc, 75.20.Hr, 78.60.-b}
\maketitle



The optical manipulation of semiconductor quantum dots (QDs) is a subject of
great interest because of its potential use to control the electronic spin for
quantum information processing \cite{ata,grei,bere} and spintronics
\cite{mac,kork,rei}. Different optical means of
manipulation \cite{rei} and detection \cite{kork} of the the spin have been proposed.
The core of the research in this area consists of optical transitions involving either neutral excitons or
trions. These states are respectively bound states of one or two electrons in the conduction band 
of the QD and a hole in the valence band, 
and can be tuned using a gate voltage $V_{g}$ \cite{ata,grei,bere,war,hog,smi,dal,klee}.

An ubiquitous aspect of the photoluminescence (PL) decay of excitons of various charges \cite{smi,dal,klee,cao,helm} or the
absorption of light creating them \cite{helm,tur,latta}, is the manifestation
of the hybridization of the orbital of the QD with a continuum of
extended states. Small QD systems can usually be described by variations of the Anderson impurity
model. For small hybridization and an odd number of
electrons in the QD, this model reduces to the Kondo model, in which the
localized spins have an exchange interaction with the spins of the electrons in the reservoir 
\cite{smi} resulting in a many-body singlet ground state \cite{hews}.

The PL spectrum that results from the decay of the trion $X^{-}$ \cite{dal,klee} 
and the neutral exciton $X^{0}$ \cite{klee} has been measured as a function
of $V_{g}$. The $X^{-}$ PL line is a consequence of an optical transition
to a state with a single electron on average in the QD. 
As a consequence of the hybridization, the PL is broad near the
limits of stability of the trion, it is asymmetric and there is a
non-monotonic blue shift. 
While simple approaches were able to explain these features \cite{dal,klee,mis} 
a fully reliable
calculation is still lacking.
For the $X^{0}$ decay, similar non-trivial
effects 
are present 
\cite{klee}. 
The creation of the 
$X^{0}$ \cite{tur} and $X^{-}$ \cite{latta} states by optical absorption were 
calculated using the numerical renormalization group (NRG), and in the $X^-$ case shows a remarkable agreement with experiment.

These transitions, because of their sudden character, are related to another
field of great interest in recent years, the dynamics of highly correlated systems,
after a quantum quench \cite{tur,latta,nor,heyl,vas,luk,anti}. 
Because initial states (IS's) and final states (FS's) have different local scattering
potentials, the spectrum should show at low temperatures a power-law behavior at the PL threshold characteristic of x-ray edge singularities \cite{helm,latta,tur,corna}.  
This is due to the Anderson orthogonality catastrophe \cite{and}, which is
another cornerstone of many-body physics, and requires sophisticated techniques for its treatment.

An important aspect of the experiment of Kleemans {\it et al.} \cite{klee} is that they 
worked in the regime of strong hybridization and studied the transition between
the $X^0$ and $X^-$ decays.
In this Letter, we calculate the PL on the whole range of gate voltages $V_g$
between the region of stability of $X^{-}$ 
and $X^{0}$, 
using NRG within the full-density matrix approach \cite{NRG1,NRG2,NRG3,DMNRG,ztrick}. 
Our results provide an explanation of recent experiments and show that a very different behavior of the 
PL spectrum under an applied magnetic field is 
expected for the $X^{0}$ and $X^{-}$ decays.

Using Fermi's golden rule, the PL intensity 
is given by

\begin{equation}
I(\omega )=\frac{2\pi }{\hbar }\sum_{if}w_{i}|\langle f|H_{LM}|i\rangle
|^{2}\delta (\hbar \omega +E_{f}-E_{i}),  \label{i1}
\end{equation}%
where $\omega $ is the PL frequency, $|i\rangle $ labels the IS, 
$|f\rangle $ denotes the possible FS's, $E_{j}$
is the energy of the state $|j\rangle $, $w_{i}$ 
is the
Boltzmann weight of the IS $|i\rangle $ and the relevant part of
light-matter interaction can be written as \cite{mis}

\begin{equation}
H_{LM}=A(d_{\uparrow }^{\dagger }p_{3/2}+d_{\downarrow }^{\dagger
}p_{-3/2})+\text{H.c.},  \label{hlm}
\end{equation}
where 
$d_{\sigma }^{\dagger }$
creates an electron at the QD with spin $\sigma $, and $p_{m}$ annihilates a
valence electron with angular momentum 3/2 and projection $m$ 
\cite{ata,note3}. 
Eq. (\ref{i1}) implies a sudden change in the dynamics of the system equivalent to a quantum quench.

\begin{figure}[tb]
\includegraphics[width=8cm]{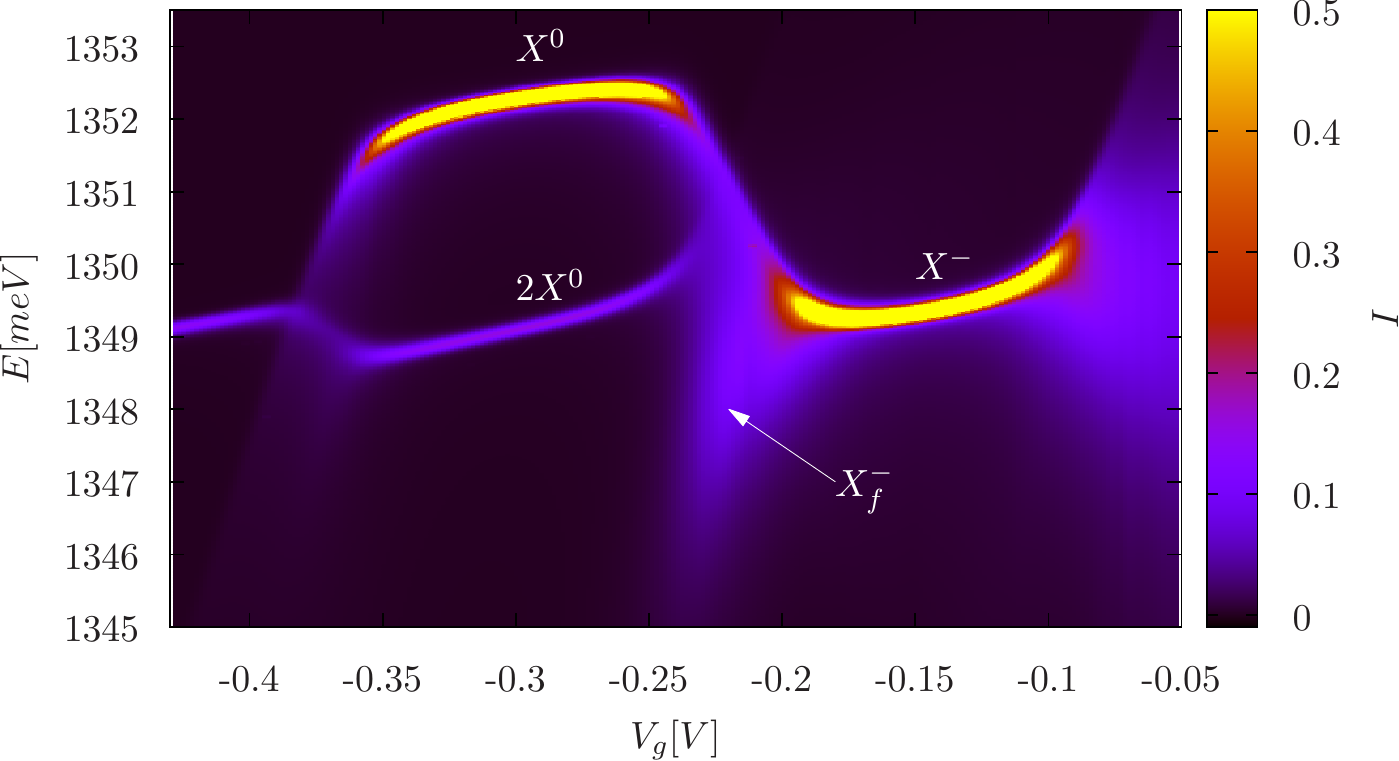}
\caption{(Color online) PL spectrum as a function of gate voltage.
The intensity is normalized by twice the maximum intensity at $V_g=-0.35V$ \cite{note3}.
Parameters are $T=1.75K$, $E_{e}^{0}=-25$ meV, $E_{h}^{0}=1398.5$ meV,
$U_{hh}=U_{eh}=U_{ee}+2 U^{\prime}_{ee} =20.97$ meV, $U^{\prime}_{ee}=3.52$ meV, 
$\Delta = 1$ meV, and $\lambda=8$.}
\label{Map_NRG}
\end{figure}

The Hamiltonian is written as

\begin{eqnarray}
	H &=&E_{e}n_d + U_{ee}n_{d \uparrow} n_{d \downarrow} 
	+\sum_{k\sigma }(V_{k}d_{\sigma }^{\dagger }c_{k\sigma }^{ }+\text{H.c.}) \notag \\
	&&+\sum_{k\sigma }\epsilon _{k}c_{k\sigma }^{\dagger }c_{k\sigma }^{ }  
	+E_{h}n_h +U_{h h}n_{h 3/2} n_{h -3/2}   \notag \\
        &&-U_{eh}n_d n_h + U^{\prime}_{ee} n_{d \uparrow} n_{d \downarrow} n_h.  \label{ham}
\end{eqnarray}
Here $h_{m}^{\dagger }\equiv p_{m}$,
$n_{d \sigma}=d_{\sigma }^{\dagger }d_{\sigma }$, 
$n_{h m}=h_{m}^{\dagger }h_{m }$, $n_d=n_{d \uparrow}+n_{d \downarrow}$, 
and $n_h=n_{h 3/2}+n_{h -3/2}$.
The first three terms correspond
to the well known Anderson impurity model. 
The remaining terms involve heavy holes,
The last term takes into account the increase
of the repulsion between electrons as their wave function contracts
after addition of holes. Its addition improves the agreement with
experiment {\em simultaneously} for the range of $V_{g}$ of both, the 
$X^-$ and the biexciton $2X^0$ decay.
Contrary to previous approaches, we consider the hybridization in both IS's and FS's. 
The on-site energies
of the QD electrons and holes change with gate voltage as $E_{e}=E_{e}^{0}-eV_{g}/\lambda$, 
and $E_{h}=E_{h}^{0}+eV_{g}/\lambda $, respectively, where $\lambda$ is the lever arm \cite{dal,klee}  

\begin{figure}[tb]
\includegraphics[width=8cm]{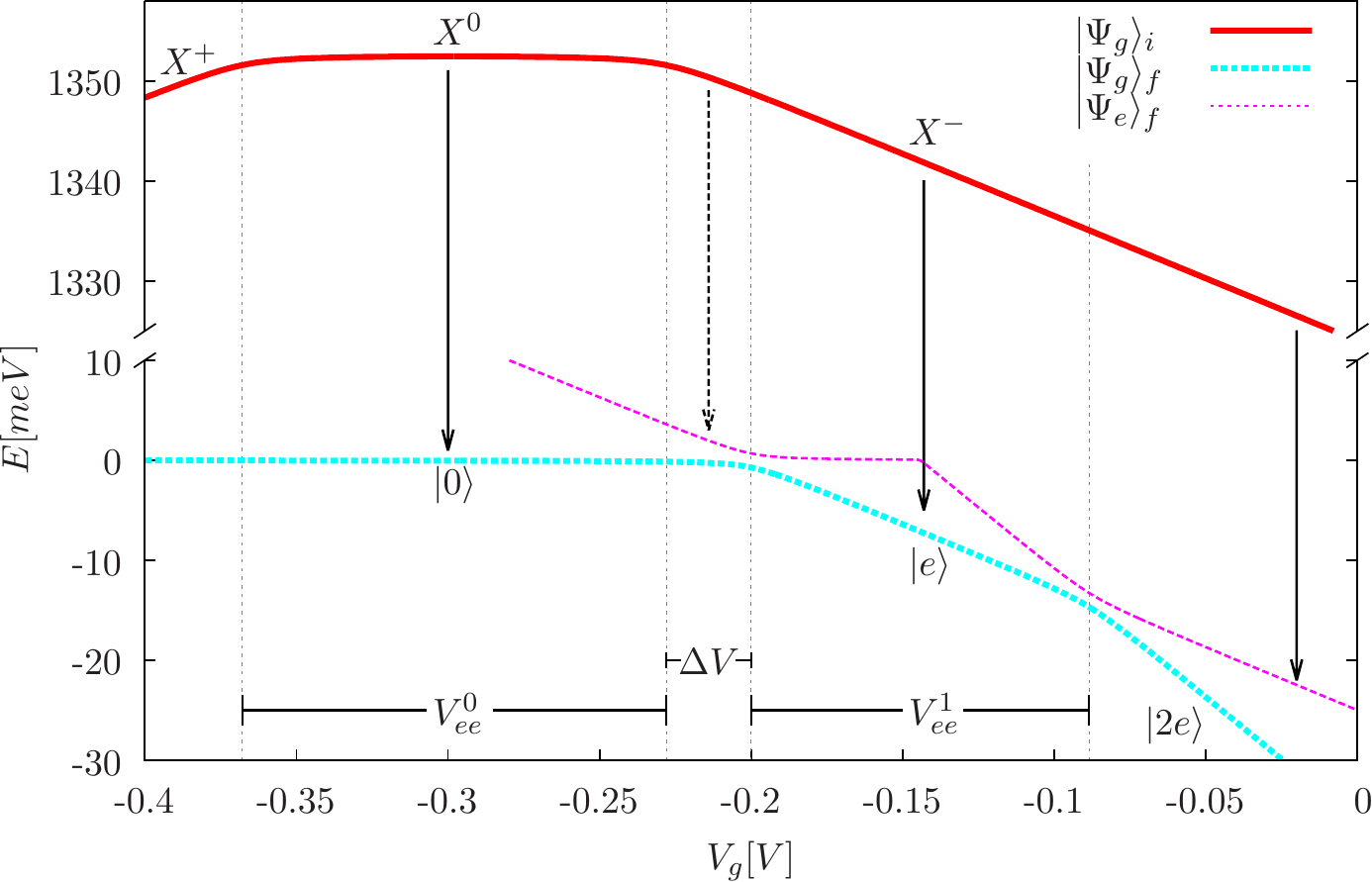}
\caption{(Color online) Most relevant energy levels of the zero-bandwidth model and the PL 
transitions (indicated with arrows).
$|\Psi_g\rangle_i$ and $|\Psi_g\rangle_f$ are the initial and final ground states, respectively.  
$|\Psi_e\rangle_f$ is an excited FS. Here $V^0_{ee}=\lambda (U_{ee}+U^{\prime}_{ee})/e$, 
$V^1_{ee}=\lambda U_{ee}/e$ and $\Delta V=\lambda(U_{eh}-U_{ee}-U^{\prime}_{ee})/e$. 
The hopping 
between the QD level and the reservoir effective level 
is $V=1$ meV. Other parameters as in Fig. \ref{Map_NRG}.}
\label{Levels}
\end{figure}

Some parameters can be determined by simple features of the experiment \cite{dal,klee}.  
For example, neglecting $V_k$, the amplitude of range of voltage in
which the PL spectrum is dominated by the decay of the trion $X^{-}$ is $V^1_{ee}=\lambda U_{ee}/e$
(see Fig. \ref{Levels}). We obtain a good agreement with experiment for the parameters indicated in Fig. \ref{Map_NRG}.
We assumed the hybridization $\Delta = \pi \sum_{k}|V_{k}|^{2}\delta ( \epsilon_F-\epsilon _{k})=1$ meV 
independent of energy with support in the range $[-D,D]$, associated with a wide reservoir band of width 
$2D=100\Delta$ symmetrically placed around the Fermi energy $\epsilon _{F}$. 
There are no particular features of the band or hybridization that can affect the results. 

We use the NRG to calculate the PL and compare with the experiments 
of Kleemans {\it et al.} \cite{klee}. We run the code two times, one in the presence of one or two holes in 
the valence  band of the QD (IS) and the other for one hole less (FS). 
Then, the density matrix for the IS's and the matrix elements entering 
Eq. (\ref{hlm}) are calculated (see e.g. Refs. \cite{corna} and \cite{latta}). 
In Fig. \ref{Map_NRG} 
we show the resulting PL spectrum as a function of $V_g$.
There are two high intensity plateaus that correspond to the decay of $X^{0}$ (lower $V_g$)
and $X^{-}$ (greater $V_g$). Near the crossover between them, the plateaus bend 
and the $X^{-}$ plateau is joined by another transition line of lower frequency and intensity,  
which corresponds to that denoted as $X_{f}^{-}$ in Ref. \cite{klee}. 
There is another plateau of lower intensity which corresponds to the decay of a biexciton, with two holes and two 
electrons to the FS $X^{0}$. This feature is denoted by $2X^{0}$ in the experiment \cite{note3}.
Except for the presence of positively charged Mahan excitons and multiple excitons which involve 
electron or hole states not included in the Hamiltonian, our results agree with those of 
Kleemans {\it et al.} \cite{klee}.

While the line shape of the PL peaks (discussed below) requires a sophisticated calculation, 
the evolution of them with $V_g$ and the transition between states of different total charge, 
can be understood qualitatively using a molecular model
in which the band width is reduced to a single electron 
reservoir state
with an energy equal to $\epsilon _{F}$ ($D \to 0$) \cite{note2}. The most relevant 
states of this model with less than two holes are represented in Fig. \ref{Levels}.
The initial ground state (IGS), before light emission, depending on the value of $V_g$ consists mainly of either 
a single hole ($X^+$) in the valence band of the QD, or $X^0$ or $X^-$, joined by regions 
with some admixture between $X^+$ and $X^0$ or $X^0$ and $X^-$ due to the hybridization.
The FS's after light emission have no hole in the valence band and, similarly to the IS's, 
mainly zero ($|0\rangle$), one ($|e\rangle$) or two electrons ($|2e\rangle$) in the 
QD. 
The arrows in Fig. \ref{Levels} correspond to the main transition in each voltage regime. 
The feature denoted as $X_{f}^{-}$ corresponds to the transition, indicated by a dashed arrow, 
from the trion to an excited state with a dot occupancy between zero and one.
The observed plateaus in the PL spectrum can be understood in the $V=0$ limit. 
The transition $X^0 \to |0\rangle$ takes place, with an emission energy 
$\hbar \omega = E_e^0+E_h^0-U_{eh}$, in the interval 
$ E_{e}^{0}-U_{eh} \leq eV_g /\lambda \leq E_{e}^{0}-U_{eh}+U_{ee}+U^{\prime}_{ee}$. 
The high $V_g$ plateau in Fig. \ref{Map_NRG}, with an energy $E_{e}^{0}+E_{h}^{0}-2U_{eh}+U_{ee}+U^{\prime}_{ee}$, 
stems from the electron-hole 
recombination of the trion $X^-\to |e\rangle$, in the range $E_{e}^{0} \leq eV_g /\lambda \leq E_{e}^{0}+U_{ee}$. 

\begin{figure}[tb]
\includegraphics[width=8cm]{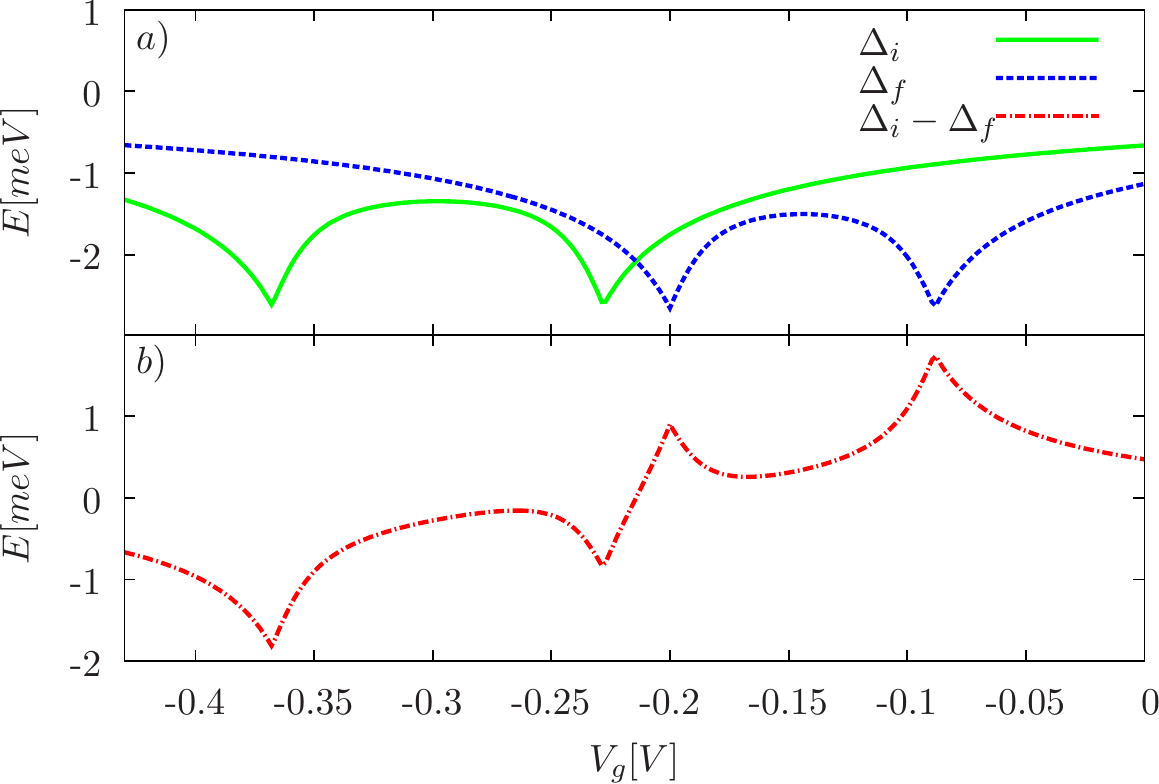}
\caption{(Color online) a) Green full line: energy gain $\Delta_i$ of the IS before photoemission
due to hybridization. Blue dashed line: energy gain $\Delta_f$ for the FS after  photoemission.
b) Change of the emitted photon energy gain $\Delta_i-\Delta_f$ due to hybridization.}
\label{delef}
\end{figure}

When the hybridization is turned on, both the IS and FS energies decrease by $\sim \Delta$.
These energy gains calculated with NRG-FDM 
and their 
difference, which gives the shift in PL frequency, are represented 
in Fig. \ref{delef}. As expected, the energy gain is larger near the intermediate valence regions.
For example, there is a dip in the energy of the FS at $V_g=\lambda E_{e}^{0}/ e = -0.2 V$ 
for which the occupancy of the FS at the dot $n_d$ is intermediate between 0 and 1. 
Another dip is clear for $V_g= \lambda(E_{e}^{0}+U_{ee})/e = -0.088 V$ for which $n_d \approx 1.5$.
Note that even in the Kondo regime $V_g \approx -0.144 V$ for which $n_d \approx 1$, the energy gain is
of order of $\Delta$ in contrast to expectations from simple approaches \cite{klee,mis}. 
For $V_g \geq \lambda E_{e}^{0}/e$,
the net effect of the hybridization is a blue shift which is more pronounced for intermediate valence occupancy
of the FS's. Similarly, in the region of the $X^0$ decay, the effect
of the hybridization is a red shift, larger near the limits of stability of $X^0$ against either
$X^+$ or $X^-$.
These shifts explain characteristic features of the PL frequency as a function of gate voltage $V_g$ observed 
in the experiment and reproduced in Fig. \ref{Map_NRG}. In particular the downward curvature in the $X^0$ region  
and an upward curvature in the $X^-$ region. A similar reasoning can be followed for the $2X^0$ feature leading
to the same qualitative behavior as for the $X^-$ decay, in agreement with the results shown in Fig. \ref{Map_NRG}. 

\begin{figure}[tb]
\includegraphics[width=8cm]{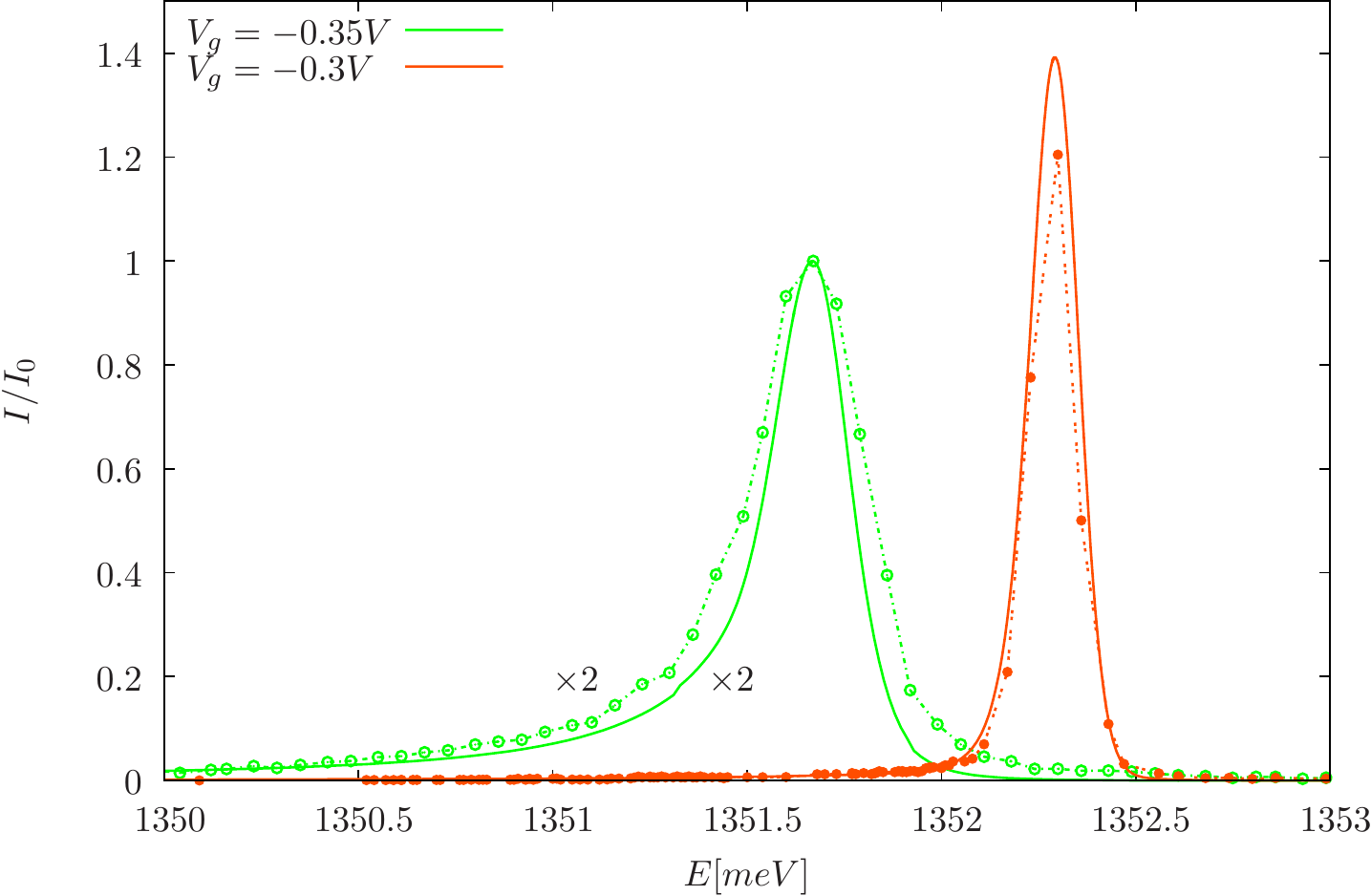}
\caption{(Color online) PL intensity close to the onset of the $X^0$ plateau (green solid line) and 
at the center of the
$X^0$ plateau (orange solid line). The orange filled and green open circles are the corresponding 
Kleemans digitized results \cite{klee}. $I_0=I(1351.6$ meV) for $V_g=-0.35V$.}
\label{Corte}
\end{figure}

\begin{figure}[tb]
\includegraphics[width=8cm]{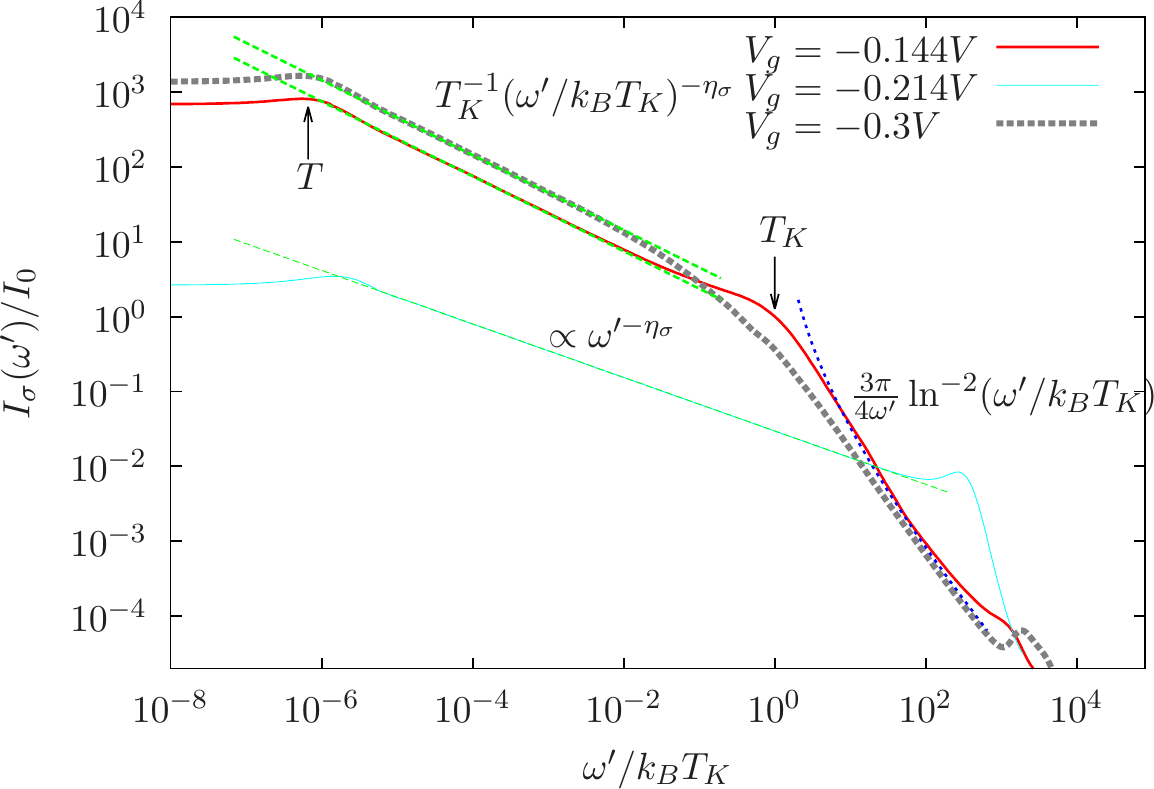}
\caption{(Color online) PL intensity as a function of the frequency $\omega^\prime=\omega_e^*-\omega$ shift from the threshold ($\omega_e^*$) in a log-log 
scale for three values of the gate voltage, $-0.144$V, $-0.214$V and $0.3$V. The threshold are $1349.74$ meV, $1350.34$ meV and $1352.54$ meV, respectively. 
The intensity is normalized by $I_0=I(k_B T_K/\hbar)$ calculated for $V_g=-0.144 V$, where $T_K\sim 0.083K$.}
\label{efx}
\end{figure}

In Fig. \ref{Corte} we show two PL line shapes that correspond to the $X^0$ decay for $V_g=-0.35V$ and $-0.3V$ and 
the respective experimental results digitized from Ref. \cite{klee}. 
The comparison is excellent.
The long tails at low energies provide evidence of the excitations produced 
by the sudden change of the dynamics of the system (quantum quench) associated with an Anderson orthogonality catastrophe. 
To discuss this point in more detail, in Fig. \ref{efx} we show the PL intensity shift from the threshold $\omega_e^*$
in a logarithmic scale, for temperatures much smaller than the Kondo temperature $T_K$ and  
three values of the gate voltage: $V_g=-0.144V$ corresponding to the Kondo regime for 
the FS in the $X^{-}$ decay, $V_g=-0.214V$ in the region between the $X^0$ and $X^-$ decays, where the 
PL frequency changes 
strongly as a function of gate voltage and both IS's and FS's are in the intermediate-valence
regime, and lastly for $V_g=-0.3V$ corresponding to the Kondo regime for 
the IS in the $X^{0}$ decay. In the first case, as the frequency is lowered, we recover the same regimes studied 
theoretically before for absorption of light creating $X^0$ \cite{tur}. 
At high frequency, the physics is dominated by the free-orbital fixed point of the NRG, with 
a dependence $I(\omega) \sim  \omega^2$. Lowering the frequency the functional form turns
to that of the local moment fixed point until the Kondo frequency $k_B T_K/\hbar$ is reached.
For $\hbar \omega < k_B T_K$, the physics enters the strong-coupling regime, and the dependence is 
the characteristic power law $\omega^{-\eta_{\sigma}}$ for the Anderson orthogonality catastrophe 
associated with the quantum quench in a Fermi liquid.
The exponent is given by \cite{tur} 
$\eta_{\sigma}=1 -\sum_{\sigma^\prime}\left(\delta_{\sigma\sigma^\prime} - \Delta n_{\sigma^\prime} \right)^2$, 
where $\Delta {n_{\sigma^\prime}}$ is the change on the local occupation of electrons with spin projection $\sigma^\prime$, 
between the FS and IS.
This divergent behavior at low frequency ceases at a frequency $k_B T/\hbar$ determined by the temperature.
For $V_g=-0.214V$, the behavior is similar, except for the absence of the local moment regime 
and the different exponent $\eta_{\sigma}$. 
For $V_g=-0.3V$ ($X^{0}$ decay), the local moment regime cannot be reached and
the PL intensity is about four times smaller at $\omega \sim k_B T_K/\hbar$ than for the 
$X^{-}$ decay. The result is a smoother curve, similar to light absorption leading 
to $X^{-}$ \cite{latta}.  

From the occupancies of the different states, 
we obtain $\eta_{\sigma}=0.4987$ for $V_g=-0.144V$, $\eta_{\sigma}=0.3571$ for $V_g=-0.214V$
and $\eta_{\sigma}=0.4983$ for $V_g=-0.3V$
The corresponding curves $\omega^{-\eta_{\sigma}}$ are also represented in Fig. \ref{efx} with 
the intensity as the only fitting parameter.

\begin{figure}[tb]
\includegraphics[width=8cm]{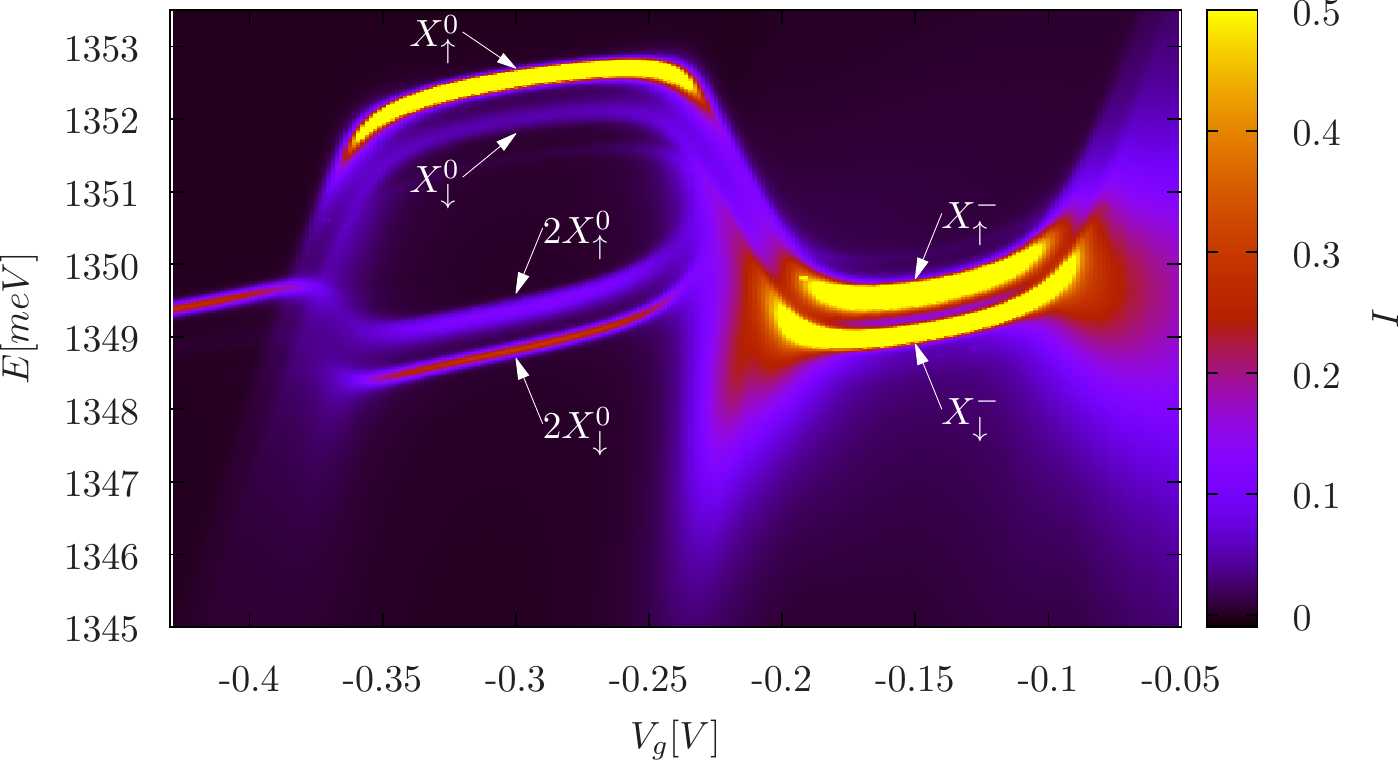}
\caption{(Color online) PL intensity in the presence of an external magnetic field producing a Zeeman splitting $0.57$ meV of the 
electronic level of the QD and a $\sim 1.05$ meV splitting for the hole level at $T=1.75K$.}
\label{fig:PLB}
\end{figure}

\begin{figure}[tb]
\includegraphics[width=8cm]{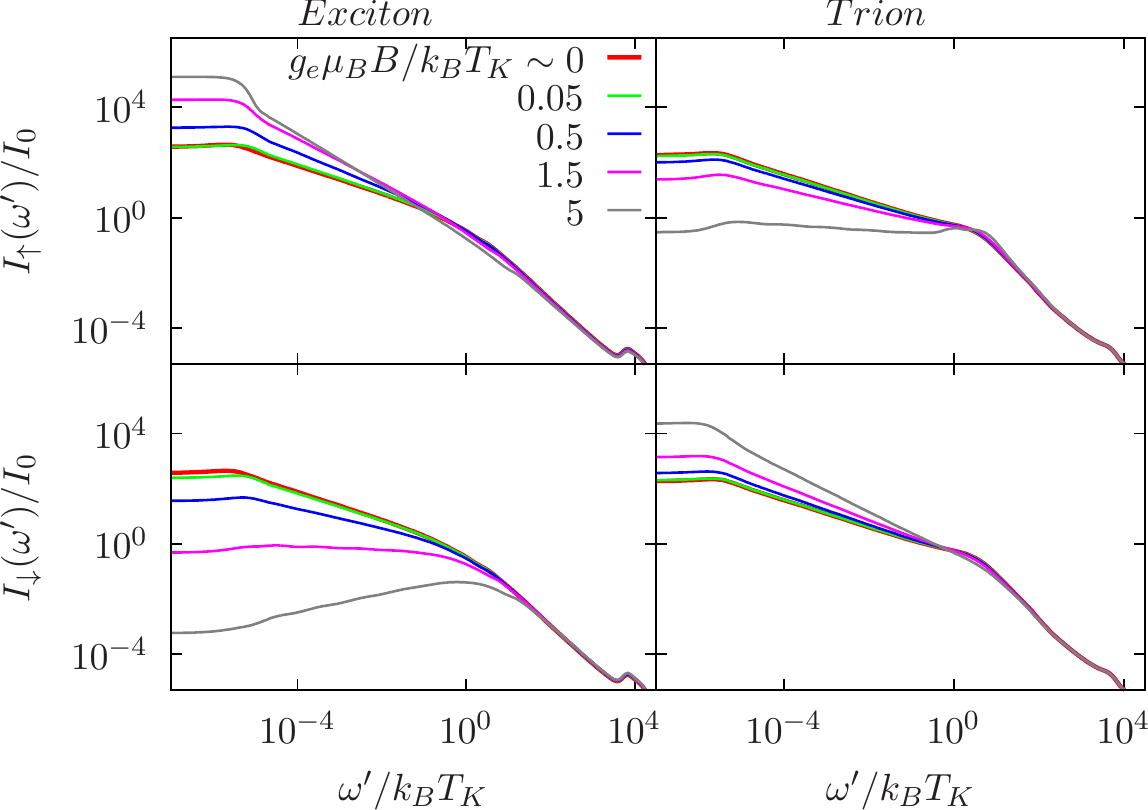}
\caption{(Color online) PL intensity as a function of the frequency shift from the threshold 
for different values of the magnetic field. 
The left side panels are in the exciton decay regime ($V_g=-0.3V$) and the right side panels are for 
trion decay ($V_g=-0.144V$).  Other parameters as in Fig. \ref{efx}.}
\label{fig:plawB}
\end{figure}

Finally, we analyze the effect of an external magnetic field $B$
\footnote{Following Ref. \cite{latta} we take the giromagnetic factors of the hole and the electron as $g_h=1.1$ and $g_e=-0.6$, respectively.}. 
As it can be seen in Fig. \ref{fig:PLB}, $B$ leads to a splitting of the PL plateaus due to the 
different giromagnetic factors for electrons and holes \cite{latta}. The different broadening of the PL peaks 
is due to the dependence on $B$ 
of $\eta_{\sigma}$
(see Fig. \ref{fig:plawB}) \cite{tur}. 
In the regime of the $X^0$ decay, the low energy plateau has a much lower intensity and is rapidly suppressed with increasing $B$. 
This is due to the reduced probability of finding an electron with a high energy spin projection in the IGS. 
For the $X^-$ decay however, the IGS has enough energy available to decay to a state with a large occupation of the high energy spin projection and both peaks have a large intensity.

In summary, using an Anderson impurity model, we can explain experimental results of PL
in a wide range of gate voltages, including those for which either IS's or FS's or both
are in the mixed valence regime. The PL transition
is an experimental realization of a quantum quench associated with another realization of Anderson 
orthogonality catastrophe, but in contrast to previous studies, in our system electron correlations are important 
in both,  IS's and FS's  states in the photon decay. We find a marked asymmetry in the PL line shapes for the 
neutral exciton and trion decays both in the hybridization induced energy shift and on the dependence with 
magnetic field. 
Absorption and emission of light are qualitatively different and the 
local moment regime can be reached only in the FS ($X^0$ absorption or $X^-$ decay).
\acknowledgements
We are supported by CONICET. This work was sponsored by PIP 112-201101-00832
of CONICET and PICT 2013-1045 of the ANPCyT.


\begin{thebibliography}{99}
\bibitem{note} In principle, Eq. (\ref{hlm}) contains another term in which 
$d_{\sigma}^{\dagger }$ is replaced by band electrons \cite{klee}. 
However this term is small and we have verified that it does not
modify substantially our results.
\bibitem{note3} To take into account the smaller probability of excitations of two holes in the experiment, 
we have multiplied the weight of the initial states with two holes by 0.1 respect to those with one hole.
\end{thebibliography}
\end{document}